\documentclass[pra,amsmath,twocolumn,showpacs]{revtex4-1}
 
\usepackage{blindtext}
\usepackage{graphicx}
\usepackage{amsmath,graphicx}
\usepackage{xcolor}
\usepackage{csquotes}
\usepackage{braket}
\def\la{{\langle}}
\def\e{\enquote}

\def\A{\mathcal A}

\def\B{\hat B}

\def\Om{\Omega}

\def\ep{\epsilon}

\def\h{\hat H}
\def\lm{\lambda}

\def\q{\quad}
\def\t{\tau}

\def\n{\\ \nonumber}
\def\G{\Psi}
\def\ra{{\rangle}}

\def\h{\hat{H}}

\def\p{{\lambda}}

\def\e{\enquote}

\def\k{{\bf k}}
\def\r{{\bf r}}

\def\I{{\text {Im}}}
\def\R{\text{Re}}

\def\q{\quad}
\def\n{\\ \nonumber}
\begin{document}
\title{Quantum reflection time and the Goos-H\"anchen effect}
\date\today
%
%
\author{D. Sokolovski$^{a,b,c,*}$}
\author{A. Uranga$^{a,d}$}
\author{Y. Caudano$^{e}$}
\affiliation{$^{a}$ Departamento de Qu\'imica-F\'isica, Universidad del Pa\' is Vasco, UPV/EHU,  48940, Leioa, Spain}
\affiliation{$^{b}$ IKERBASQUE, Basque Foundation for Science, E-48011 Bilbao, Spain}
\affiliation{$^{c}$ EHU Quantum Center, Universidad del Pa\' is Vasco, UPV/EHU, 48940 Leioa, Spain}
\affiliation{$^{d}$ Basque Center for Applied Mathematics (BCAM), 48009, Bilbao, Spain}
\affiliation{$^{e}$ 
Department of Physics, Namur Institute for Complex Systems (naXys) and Namur Institute of Structured Matter (NISM), University of Namur, Rue de Bruxelles 61, 5000 Namur, Belgium}
\email{dgsokol15@gmail.com}
\begin{abstract}
\noindent
We explore the analogy between following the motion  
of a reflected wave packet, and a quantum measurement
of the spatial delay imposed on  the  particle  by the scattering  potential.
It is shown that converting such delays into temporal durations can 
lead to \e{negative times}, and best be avoided. 
It is also demonstrated that a \e{soft} potential can be replaced 
by  superposition of hard walls. This representation is used 
for calculating the Goos-H\"anchen shift of a Gaussian beam
incident on a potential step or a barrier at an oblique angle. 

 \date\today
\end{abstract}

%
%
\maketitle
\section{Introduction}
The Goos-H\"anchen (GH) effect, first observed in 1947 for electromagnetic waves \cite{GH0}, consists in lateral displacement
of a beam reflected off a plane surface. The effect is not unique to electromagnetism,  and can occur in a variety 
of wave phenomena, including quantum scattering, see for example  \cite{GH2}-\cite{GH4}. The displacement is often related 
with the time delay, associated with the  scattering process \cite{GH1}, and can be seen as a way  of measuring it. 
However, there is still little consensus about the definition of the quantum mechanical time delay. 
Much attention has been paid to the \e{tunnelling time problem} \cite{REV1}-\cite{REV2} of defining the duration a tunnelling particle 
is supposed to spend in the barrier region. Recently we argued that the answer is provided by the measurement theory, 
in a uniquely quantum way \cite{DS1},  \cite{DS2}, \cite{DS3}. According to the Uncertainty Principle \cite{FeynL}, to know the duration of interest one needs to 
destroy the interference, and that destroys tunnelling. Conversely, one can preserve tunnelling, but with interference intact 
the duration of interest must remain indeterminate \cite{DS1}, \cite{DS2}. The same should be true in quantum reflection, 
where one would expect  similar non-classical  behaviour for a particle repulsed by a potential well, or reflected above 
a potential barrier. 
\newline
In this work, we revisit the quantum Goos-H\"anchen effect in order to establish whether determination of the GH shift 
does amount to measuring the time spent by the particle in the scattering potential. 
We will answer the question by the negative, and proceed to determine what is being measured, and to what accuracy. 
The  paper is organised as follows. In Section II, we will rewrite the reflected wave packet (WP) in terms 
of the freely propagating counterpart, and a number of its mirror images.  In section III we exploit the analogy with 
a von Neumann quantum measurement performed on a pre- and post-selected quantum system, 
and identify the measured quantity. In Sect. IV we show that in the classical limit one does 
determine  the duration spent by the classical trajectory in the scattering potential. 
In Sect. V we attempt to extend the idea to the fully quantum case, and fail. 
In Sect. VI we obtain the reflection \e{phase time} and discuss its properties. 
Sects. VII and VIII offer two simple examples of what has been said so far. 
In Sect. IX we abandon the attempts to define a meaningful \e{reflection time}, 
and replace instead a soft potential barrier by a superposition of virtual 
hard walls. In Sect. X, we use this representation to reconstruct the reflected 
beam, and evaluate the GH shift in terms of the spatial shifts experienced by 
the reflected particle. Section XI contains our conclusions. 
\section{Reflection in the momentum and  coordinate representations}
We consider a potential potential $V(x)$, 
contained in the region $0<x<d$, and a wave packet (WP) state
$|\Psi(0)\ra$, $\la x|\Psi(0)\ra\equiv 0$ for $x>0$. 
\begin{figure}[h]
\includegraphics[angle=0, width=0.48\textwidth]{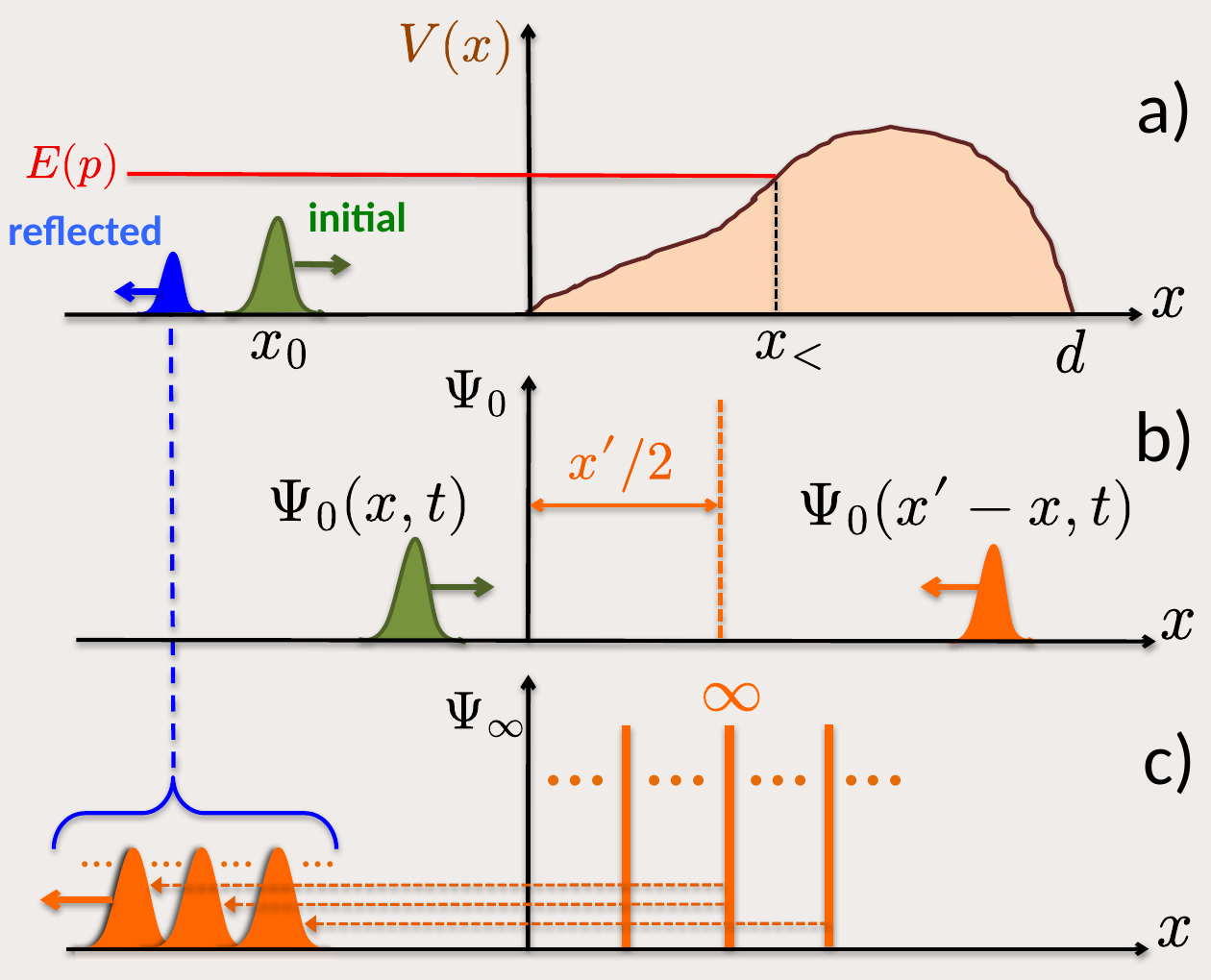}
\caption {a) A wave packet with mean momentum $p$, incident on a potential barrier $V(x)$, 
is partly reflected. Its mean energy  is lower than the barrier height, 
and $x_<$ denotes the classical turning point $V(x_<) = E(p)$.
b) Freely propagating WP and its reflection is a mirror  placed at $x=x'/2$.
c) The reflected WP shown in (a) is a superposition of the WPs reflected 
by a variety of  infinite walls.}
\label{Fig:1}
\end{figure}
A convenient choice is a Gaussian WP moving towards the barrier (see below).
  The initial state at $t=0$ can be expanded in plane waves,
 (we use $\hbar=1$) 
 \begin{align} \label{b0}
\Psi(x,0)=\int_{-\infty}^\infty dk \la x|k\ra\la k|\Psi(0)\ra, 
\end{align}
where $\la x|k\ra =(2\pi)^{-1/2}\exp(ikx)$.
Time evolution of the WP is found by expanding it in the complete set of scattering states, incident on the barrier from the left and from the right (see Appendix A).
Thus, for  $t>0$ the wave function in the left half-space $x<0$ is given by
 \begin{align} \label{b1}
&\G(x,t) = \G_0(x,t)+  \\
&\int_{-\infty}^\infty \frac{dk}{\sqrt{2 \pi}}  R(k) \la k|\Psi(0)\ra \exp [-ikx-iE(k)t],\ \ [x<0],
\nonumber
\end{align}
where
\begin{align} \label{b2}
\G_0(x,t) =  \frac{1}{\sqrt{2 \pi}}\int_{-\infty}^\infty dk \la k|\Psi(0)\ra \exp [ikx-iE(k)t],
\end{align}
is the freely evolving initial state, 
and $R(k)$ is the reflection amplitude for a particle with momentum $k$, incident on the potential from the left.
The function $R(k)$ is known to be analytic in the complex $k$-plane, with poles in its lower half, and also on the 
positive imaginary axis, if $V(x)$ supports bound states \cite{BZP}. 
Note that  Eq.(\ref{b1}) is valid even when the initial state (\ref{b0}) contains negative momenta, in which case one uses
$R(-|k|)=R^*(|k|)$ (see Appendix A). 
\newline
One can transform Eq.(\ref{b1}) to the coordinate representation by
inserting the identity 
 \begin{align} 
R(k) &= \int_{-\infty}^\infty dx' \exp(ikx') \xi(x') \label{b3b}\\
\xi(x') &=(2\pi)^{-1}  \int_{-\infty}^\infty dk' \exp(-ik'x') R(k') \label{b2a}
\end{align}
into Eq.(\ref{b1}). This yields 
 \begin{eqnarray} \label{b3}
\G(x,t) = \G_0(x,t) +\int_{-\infty} ^\infty dx'  \G_0(x'-x,t)\xi (x').
\end{eqnarray}
where $\G_0(x'-x,t)$ is the image of the freely propagated WP (\ref {b2}) in a mirror placed at  $x=x'/2$ (see Fig.\ref{Fig:1}b). 
Each image enters the superposition (\ref{b3}) with a weight $\xi (x')$,  the Fourier 
transform of the reflection amplitude with respect to particle's momentum. 
For a  potential $V(x)$, such that $R(k)\to 0$ as $|k|\to\infty$, $\xi(x')$ is a smooth function of $x'$, 
and, in the absence of bound states, one has
  \begin{eqnarray} \label{b3a}
\xi(x')\equiv 0 \q \text {for} \q x'<0.
\end{eqnarray}
\newline 
The evolution of the wave function in the left half-space can now be visualised as follows.
At first, the WP localised far to the left, moves freely towards the scatterer, and the wave function 
is given by the first term in Eq.(\ref{b1}), $\Psi(x,t) \approx \G_0(x,t)$. 
 For a while its mirror images $ \G_0(x'-x,t)$,  moving  in the opposite direction, remain
\e{off screen}
in the right half-space $x>0$. 
As the free WP crosses the origin,  
and disappears from the {screen},  its mirror images enter the region $x < 0$, replacing there 
$\G_0(x,t)$. Scattering is finished when the left half-space contains only
 the reflected (R) part of the wave function, 
  \begin{eqnarray} \label{b4}
\G(x,t){\xrightarrow[ t \to \infty  ] {}} \int_{-\infty} ^\infty dx' \xi (x')  \G_0(x'-x,t)\equiv \G_R(x,t).\q \q
\end{eqnarray}
Needless to say, the representation (\ref{b3}) has no physical significance  in the {right} region $x>0$,
but is exact for any $x<0$. 
\section{Analogy  with a quantum measurement}
Let the initial state of a particle of mass $m$ be a Gaussian WP 
with a mean momentum $p=mv>0$, and coordinate width $\Delta x$, 
placed at $x=x_0<0$. 
Throughout the paper we will neglect the WP's  spreading,
so the free state in Eq.(\ref{b2}) propagates without distortion,
 (see Appendix B),  
and we have
  \begin{eqnarray} \label{e1}
\Psi_0(x,t)
\approx G_0(x-x_0-vt)\exp\left[i p (x-x_0)- i E(p) t\right], 
\end{eqnarray}
where
 \begin{eqnarray} \label{e1a}
G_0(x)=\left (\frac{2}{\pi\Delta x^2 }\right)^{1/4}\exp\left ( -\frac{x^2}{\Delta x^2}\right). 
\end{eqnarray}
Reflecting $\Psi_0(x,t)$  about the point $x'/2$ requires changing $x\to x'-x$, and for the reflected state (\ref{b4}) we find
  \begin{align} \label{e3}
\G_R(x,t) &\approx\exp[-i p (x+x_0)- i E(p) t]\times\\
 \int_{-\infty}^\infty& dx' G_0(x'-X(t)) \exp(i p x')\xi(x'), \q [x<0],\nonumber
\end{align}
where we introduced a shorthand 
 \begin{eqnarray} \label{e3a}
 X(t)\equiv x +x_0+vt. 
 \end{eqnarray}
Similarity with a quantum measurement made on a pre- and post-selected quantum system is now evident.
The amplitude of finding the particle at $x$, $\G_R(x,t )$, matches that of a von Neumann pointer \cite{vN} of a mass
$m$ and momentum $-p$, set up to determine the manner in which a post-selected system reaches its final state [see Eq.(\ref{ac7}) of the Appendix D].
As with the von Neumann pointer, the amplitude $\G_R(x,t )$
can be seen as a result of interference between 
various alternatives, each labelled by $x'$.
The  experimenter, who finds the reflected particle in $x$ at $t$, knows that $x'$ (whatever it may be) 
lies in the interval 
$X(t)-\Delta x \lesssim x' \lesssim X(t)+\Delta x$, since only the values within this range
 contribute to the corresponding probability $|\G_R(x,t )|^2$.
Destruction of interference between the alternatives and, therefore, the accuracy with which $x'$ is determined, 
is the greater the smaller is the WP's width $\Delta x$. 
\newline
The peculiarity of the situation is that, unlike in the von Neumann example, no other degree of freedom is involved, 
so the particle appears to \e{measure itself}. But what precisely is being measured?
 Rewriting the reflected plane wave as 
  \begin{eqnarray} \label{e5}
R(p) \frac{\exp(-ipx)}{\sqrt{2 \pi}}  = \int_{-\infty}^\infty \frac{dx'}{\sqrt{2 \pi}} \exp [-ip(x-x')] \xi(x'),\q 
\end{eqnarray}
we note that $x'$ represents a {\it spatial displacement}, or {\it shift} with which 
 the reflected particle  leaves  the scatterer, 
with $\xi(x')$ giving the corresponding probability amplitude.  
\newline
The usual dilemma 
 is now evident. 
In the von Neumann example of Appendix D,  interaction with the pointer alters the likelihood of the post-selection. 
 Similarly, the reflection probability $P_R(t) \equiv \int dx |\G_R(x,t)|^2$ depends on the WP's width $\Delta x$.  
An accurate measurement,  $\Delta x \to 0$, destroys reflection, $P_R(t)\to 0$, since most of the WP,  broad in the momentum space, 
and is now transmitted.  Conversely, the choice of a nearly monochromatic WP, $\Delta x \to \infty$,  preserves $P_R(t)\to |R(p)|^2$ but fails to distinguish 
between different shifts $x'$. 
A similar claim that, in a double slit experiment, one 
cannot both know the slit chosen by the particle and retain the  interference pattern on the screen 
is often used to illustrate the Uncertainty Principle \cite{FeynL}.
\newline
It is instructive to see how the notion of reflection time occurs in the classical limit of the just described \e{measurement}.
\section {(Semi)classical {reflection time}}
Consider the semiclassical limit of Eq.(\ref{e3}) in the  case of the barrier shown in Fig.\ref{Fig:1}a. For energies less than the barrier
height the reflection amplitude is given by  \cite{Land}
 \begin{eqnarray} \label{f1}
R(k)\approx -i \exp\left [2i \int_0^{x_<(k)}q(x,k)dx\right ],
\end{eqnarray}
where $q(x,k)=\sqrt{k^2-2mV(x)}$ is the local momentum, and $x_<$ is the turning point, where $q(x_<(k),k)=0$
(see Fig.\ref{Fig:1}a). The integral over $k$ in Eq.(\ref{b1}) contains rapidly oscillating exponentials 
and can be evaluated 
by the stationary phase method \cite{St}. Integral over $dx'$ in Eq.(\ref{b4}) is also oscillatory and, as is easy to check, 
has a stationary point at $x'_s(p)$, 
  \begin{eqnarray} \label{f2}
\frac {mx_s'(p)}{p}=2\int_0^{x_<(p)}\frac{mdx}{q(x,p)}.
\end{eqnarray}
As a result a single envelope is selected in Eq.(\ref{e3}),
  \begin{eqnarray} \label{f4}
\G_R(x,t) \approx  \exp[-i p (x+x_0) - i E(p) t]\q\q\q\q\q\q\q \n G_0(x_s'(p)-X(t)).\q
\end{eqnarray}
We note that Eq.(\ref{f2}) equates the time it takes the free particle to travel 
a distance $x'$ to the actual time of travel from the origin $x=0$  to the turning point $x_<$ and back. 
The classical time 
measurement  now looks like this. The experimenter prepares a classical particle 
with momentum $p$ at $x_0<0$ to the left of the barrier, and waits until it bounces off it.
Then he determines its position $x$ at time $t$, calculates $x'_s=X(t)=x+x_0+vt$, and finally
finds  the duration spent in potential, 
  \begin{eqnarray} \label{f5a}
\t=\frac{mx_s'}{p} \equiv \frac{x'_s}{v}.
\end{eqnarray}
\newline 
For a semiclassical particle, prepared in a Gaussian state, a convenient reference point is 
the centre of mass of the reflected WP (\ref{f4}),  
  \begin{eqnarray} \label{f5}
\la x \ra\equiv \frac{\int_{-\infty}^0 dx\, x\, |\G_R(x,t)|^2}{ \int_{-\infty}^0 dx| \G_R(x,t)|^2}
\end{eqnarray}
from which one obtains 
  \begin{eqnarray} \label{f6}
\la \tau\ra \equiv \frac{\la x\ra}{v}+\frac{x_0}{v} +t.
\end{eqnarray}
The classical experiment cannot, however, determine such a duration for a reflected particle
with energy higher than the barrier top, or for a particle reflected  by a potential well.
Such events are, however, possible  in the quantum case which we will consider next.
\section {Quantum \e{reflection time}}
One way to extend  the classical result  (\ref{f5a}) to a fully quantum case is to
associate a duration $\t=x'/v$ with each spatial  shift $x'$ in Eq.(\ref{e3}). 
   Now having found the reflected particle in $x$ at $t$, one can assume 
  that it had spent in the potential a duration $X(t)/v-\Delta x/v \lesssim \t \lesssim X(t)/v+\Delta x/v$.
  The danger, as will be shown shortly,  is that one will  encounter 
  negative durations whenever $V(x)$ provides negative shifts, 
  $\xi(x'<0)\ne 0$. We will proceed nonetheless.  
\newline
For an  accurate measurement, one requires $\Delta x \to 0$, and 
since $\xi(\t)$ is a smooth function,
one finds ($x<0$),
  \begin{eqnarray} \label{g2}
  |\G_R(x,t)|^2 \approx 
\left | \int_{-\infty} ^\infty dx' G_{0}(x')\right |^2 \left |\xi(x+ vt+x_0)\right  |^2. \q
\end{eqnarray}
With interference between the alternatives destroyed, a single duration $\t(x)=t+\frac{x}{v}+\frac{x_0}{v}$ can be ascribed to a particle found in $x$.
Averaging over all positions yields the mean \e{reflection time} (\ref{f6})
   \begin{eqnarray} \label{g3}
\la  \t \ra 
{\xrightarrow[ \Delta x \to 0  ] {}}
\frac{1}{v}
\frac{\int_{-\infty}^\infty dx\, x\, |\xi(x+v t +x_0)|^2}{\int_{-\infty}^\infty dx\, |\xi(x + v t + x_0)|^2}.
\end{eqnarray}
The danger here, we repeat, is that, since $\xi(x')$ may not  vanish for $x'<0$, the average (\ref{g3}) can, in principle, turn negative, and no longer 
represent a meaningful duration. 
\newline
We note also that since $G_0(x) {\xrightarrow[ \Delta x \to 0  ] {}}(2\pi\Delta x^2)^{1/4}\delta(x)$, the first integral in Eq.(\ref{g2}) behaves as $\sim \Delta x ^{1/2}$, and $|\G_R(x,t)|^2\to 0$, 
because most of the momenta now go above the barrier (see also Sect.III). 
\newline
Next we consider the case of a highly inaccurate measurement, $\Delta x \to \infty$,
where all the alternatives, be they spatial shifts $x'$, or times $x'/v$, interfere 
to produce the probability $|\G_R(x,t)|^2$ of finding the reflected particle in $x$. 
\section { \e{Phase time} for reflection}
As the initial wave packet becomes broader, $\Delta x \to \infty$, 
$G_{0}(x'-X(t))$
 in Eq.(\ref{b4})
varies little across the range of the $x'$s contained in the $\xi_\t(\t)$, and can be approximated 
by the first two terms of its Taylor series, 
$G_{0}(x'-X(t))\approx G_{0}(X(t))- x'\partial_{x}G_{0}(X(t))$.
Inserting this into (\ref{e3}),
  yields
  \begin{eqnarray} \label{h1}
\G_R(x,t) \approx\exp[-i p (x+x_0)-i E(p) t] \times\q\q\q\q\q\q\q \n
G_{0}(X(t))\int dx' \eta(p,x') -  \partial_{x}G_{0}(X(t))\int dx' x'  \eta(p,x')\q
\end{eqnarray}
where we wrote 
$\eta(p,x')\equiv \exp(ipx')\xi(x')$. 
Inserting (\ref{h1}) into (\ref{f5}), and using (\ref{f6}) yields the \e{phase time}
   \begin{eqnarray} \label{h2}
   \la  \t \ra 
{\xrightarrow[ \Delta x \to \infty  ] {}}
\frac{1}{v}\R\left[ \frac{\int_{-\infty}^\infty dx'  x'\eta(p,x')}{\int_{-\infty}^\infty dx' \eta(p,x')}\right ]\n
=\frac{1}{v}\frac{\partial \phi_R(p)}{\partial k}
\equiv \t_{ph},\q\q
 \end{eqnarray}
where $\phi_R(p)$ is the phase of the reflection coefficient, $R(p)=|R(p)|\exp[i \phi_R(p)]$.
The phase time (\ref{h2})
is often thought  to represent the 
duration a particle with momentum $p$ spends in the reflecting potential.
[In quantum measurements, the complex quantity in the brackets is known as \e{weak value of $x'$}\cite{W1}, \cite{W2}.
It can be shown that, as in the case of transmission \cite{DS3}, 
its imaginary part determines the small change in the reflected particle's mean momentum, 
which occurs because higher momenta are transmitted more easily, and are less likely to be reflected (see also \cite{GH4}).]
\newline
However, one needs to be even more cautious. 
Since complex valued $\eta(p,x')$ may change sign, 
$\t_{ph}$ can be negative,  even if all phase shifts in Eq.(\ref{e5}) are positive  
 and $\xi(x')$ vanishes  for $x' <0$ (for an example see \cite{Neg}). 
\newline
Simple illustrations will be given next.
\section{Zero-range potentials. Negative  \e{reflection times}}
For a particle 
scattered by a  zero-range  potential  $V(x) =\Om \delta(x)$, 
where $\delta (x)$ is the Dirac delta,
the reflection amplitude is well known to be
  \begin{eqnarray} \label{d2}
R(k)= \frac{-im\Om}{k+im\Om}.
\end{eqnarray}
Thus, we have
  \begin{eqnarray} \label{d3_a}
\xi(x') = m \vert\Om\vert \exp(- m \vert\Om  x' \vert)\, \theta(\pm x'),
\end{eqnarray}
where the upper and lower signs are  for a barrier, $\Om>0$, and a well $\Om<0$, respectively.
\newline
For a $\delta$-barrier, both the average (\ref{g3}) and the phase time (\ref{h2}) are positive
\begin{align} \label{d3_b}
\la  \t \ra {\xrightarrow[ \Delta x \to 0  ] {}}\frac{1}{2\vert\Om\vert v}>0,\q\
 \tau_{ph}(p)=v^{-1}\frac{m \vert\Om\vert}{p^2+m^2 \Om^2}>0. \q
\end{align}
For a $\delta$-well \cite{FOOT} both $\la  \t \ra$ and $ \tau_{ph}$  change sign and become negative, 
\begin{align} \label{d4}
\la  \t \ra {\xrightarrow[ \Delta x \to 0  ] {}}\frac{-1}{2\vert\Om\vert v}<0,\q
 \tau_{ph}(p)=v^{-1}\frac{-m \vert\Om\vert}{p^2+m^2 \Om^2}<0. \q
\end{align}
The quantities in Eqs.(\ref{d4})  cannot  be interpreted as durations the particle spends in the right half-space.
The fault is with conversion of the phase shifts in Eq.(\ref{e5}) into 
temporal intervals, which yields a negative time $\t$ whenever $x'<0$. 
Finally, the
reflection probability 
 \begin{eqnarray} \label{d3_c}
P_R
{\xrightarrow[ \Delta x \to 0  ] {}}\sqrt{2\pi} \Delta x\, m |\Om|/2\q\q
\end{eqnarray}
vanishes, as expected,  for a WP narrow in the coordinate space.
\section{A step potential. Infinite phase time}
\begin{figure}[h]
\includegraphics[angle=0,width=0.48\textwidth]{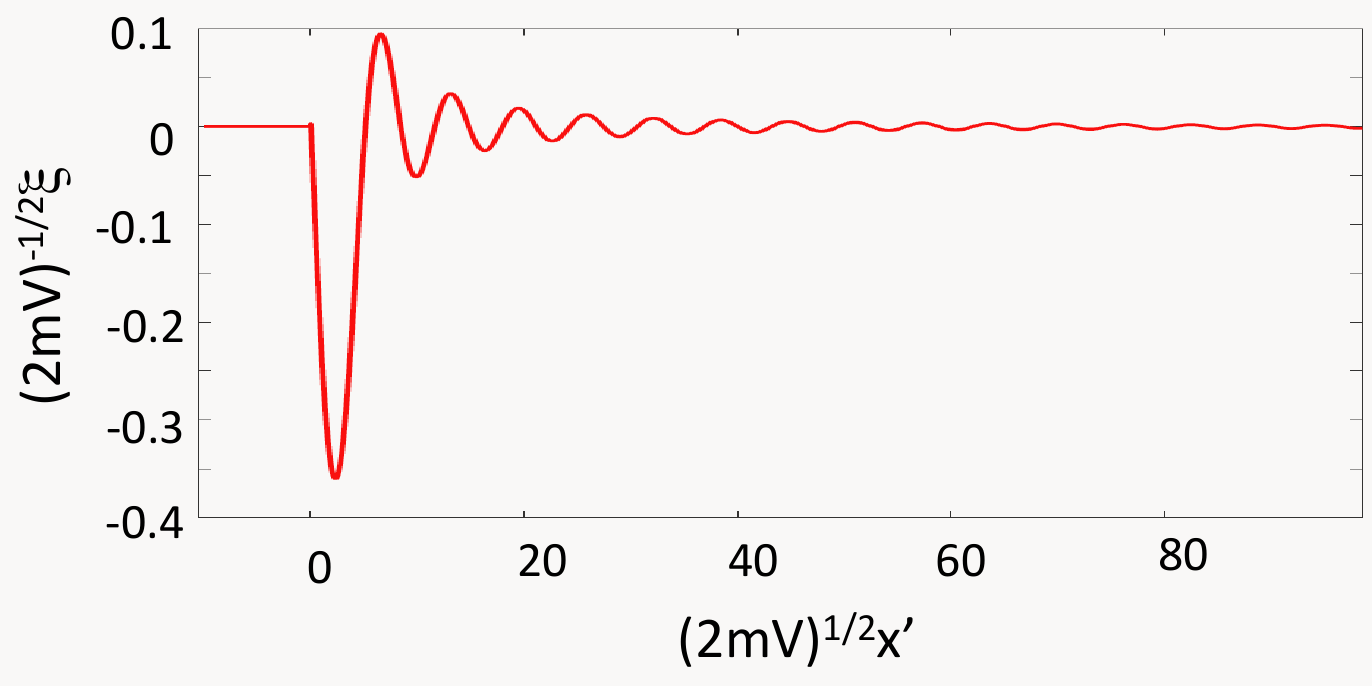}
\caption {The amplitude distribution $\xi(x')$ for a step potential (\ref{m1}), as given by (\ref{ad3text}).}
\label{Fig:2}
\end{figure}

Consider next a step potential
  \begin{eqnarray} \label{m1}
V(x)=V\theta(x), 
\end{eqnarray}
for which the reflection amplitude is easily found to be 
 \begin{eqnarray} \label{m2YC}
R(k,q(k))=\frac{k-q}{k+q},  
\end{eqnarray}
where  $q(k)=\sqrt{k^2-2mV}$.
Evaluation of the integral (\ref{b2a}) yields (for details see Appendix E)
  \begin{eqnarray} \label{ad3text}
\xi(x') =
 - \frac{2}{x'}\, J_2(\sqrt{2mV} x')\, \theta(x')
\end{eqnarray}
where  $J_2(z)$ is the Bessel function of the first kind of order two \cite{Abram} (see Fig.\ref{Fig:2}).
It is readily seen that $\int_0^\infty dx' \xi(x')=-1$ and,  as the height of the step 
increases,   $\xi(x'){\xrightarrow[ V \to \infty  ] {}}{-\delta(x')}$,  the reflection amplitude tends to that of a hard wall, 
 \begin{eqnarray} \label{m2}
R(p)=\int_0^\infty dx' \exp(ipx')\xi(x')
{ \xrightarrow[ V \to \infty  ] {}}-1. 
\end{eqnarray}
\begin{figure}[h]
\includegraphics[angle=0,width=0.48\textwidth]{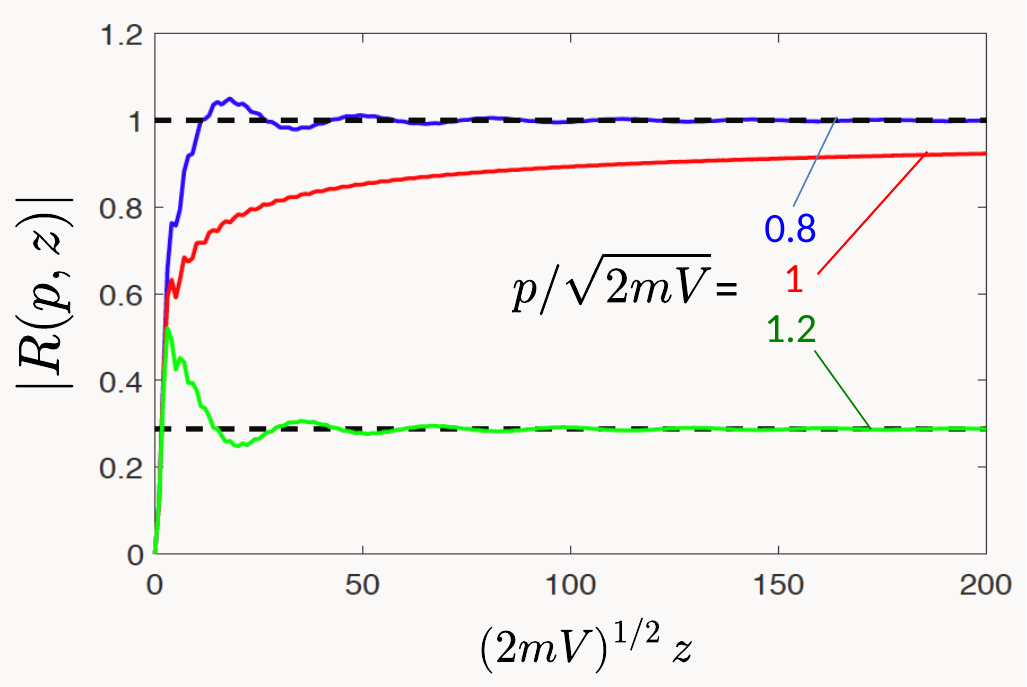}
\caption {Integral $R(p,z) \equiv \int_0^z dx' exp(ipx')\xi(x')$ converges to 
the reflection amplitude $R(p)$(dashed) as $z\to \infty$. Convergence is slower
for energies close to the step's height, $p\approx \sqrt{2mV}$ }
\label{Fig:3}
\end{figure}
To understand the behaviour of the phase time $\t_{ph}$, we note first that the integrand in the denominator of  Eq.(\ref{h2}) contains  the term $\exp[i(p-\sqrt{2mV})x']/x'^{3/2}$, 
and converges, as illustrated in Fig.\ref{Fig:3}. The integrand in the numerator of Eq.(\ref{h2}), fall off as  $\sim 1/x'^{1/2}$, and
 \begin{eqnarray} \label{k0YC}
\t_{ph}(p) =
 \begin{cases} 
\frac{2m}{p\sqrt{2mV-p^2}}& \text{ for}\q  0\leq p \leq  \sqrt{2mV}\\
0& \text {for}\q    p >  \sqrt{2mV},\q
\end{cases}
\end{eqnarray}
becomes infinite as $p\to \sqrt{2mV}$. 
The fault is with the approximation (\ref{h1}), 
without which both $\la x \ra$ in Eq.(\ref{f5}) and the $\la \t\ra$ defined by 
Eq.(\ref{g3})
 remain finite, as shown in Fig.\ref{Fig:4} (see also  \cite{GH4}).
\begin{figure}[h]
\includegraphics[angle=0,width=0.48\textwidth]{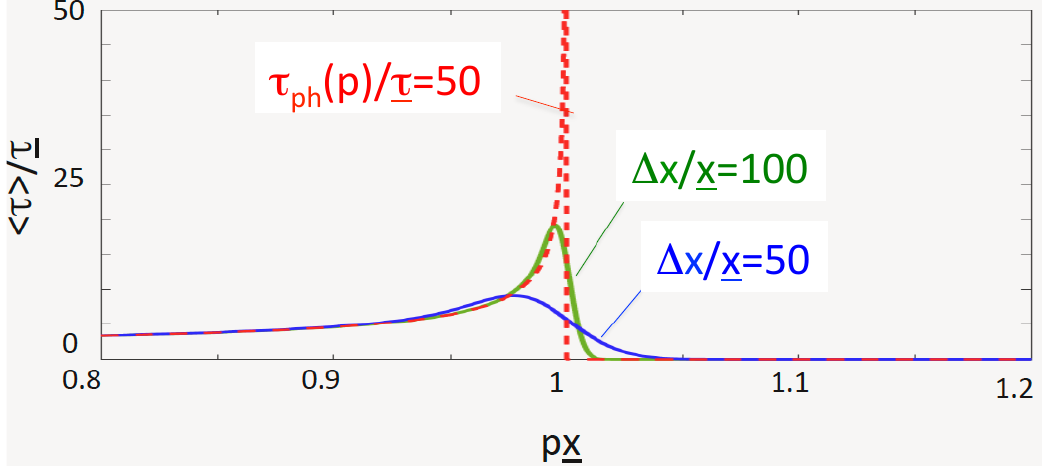}
\caption {The centre-of-mass delay for a WP incident on the potential step (\ref{m1}) for two 
coordinate widths. Also shown by the dashed line is the phase time (\ref{h2}), which diverges
when $E(p)=V$.
The units are $\underline x=1/\sqrt{2mV}$ and $\underline \t=p^{-1}(2V/m)^{-1/2}$. }
\label{Fig:4}
\end{figure}
\section{Soft barrier as a superposition of hard walls}
Having abandoned attempts to describe quantum reflection in terms of a \e{reflection time}, 
we look at the problem from a different angle.
In Eq.(\ref{b4}) the free state  $\Psi_0(x'-x,t)$ differs only by sign from the state $\Psi_R^\infty (x,t|a)$, reflected 
by an infinite hard wall placed at $a=x'/2$, and we can write
 \begin{eqnarray} \label{c3}
\G_R(x,t ) = \int_{0} ^\infty da\, \xi_a(a)  \G^\infty_R (x,t|a), \q [x<0].
\end{eqnarray}
where $ \xi _a(a)=-2\; \xi (2a)$, and
  \begin{eqnarray} \label{c4}
\G^\infty_R (x,t|a)=\q\q\q\q\q\q\q\q\q\q\q\q\n
- \frac{1}{\sqrt{2 \pi}} \int_{-\infty}^\infty dk \la k|\Psi(0)\ra \exp[ik(2a-x)-iE(k)t]\q
\end{eqnarray}
As long as only the reflected part of the WP is of interest, one can replace the actual potential $V(x)$ by a superposition of hard walls,
$V_\infty(x|a) = 0$ for $x<a$, and $\infty$ otherwise. The reflected WP is now given by the superposition
of the states reflected by all infinite walls, with appropriate amplitude weights $\xi_a (a)$. 
Now $x'=0$ corresponds to reflection by a wall placed at the origin, in which case
the particle does not enter the region $x>0$. For a positive $x'>0$ the wall is placed inside the $x>0$ region, 
so the centre of the  WP, reflected from it,  lags behind $\G^\infty_R (x,t|0)$. The opposite is true for $x'<0$.
\newline 
If the potential supports bound states, some of the walls will have to be placed in the $x<0$ region, suggesting reflections 
occurring even before the particle impacts the potential [cf. Eq.(\ref{d4})]. 
However, for a pure barrier, having no poles in the upper half of the complex $k$-plane, we have  
 \begin{eqnarray} \label{c0}
\xi_a(a)\equiv 0, \q \text{for}  \q a<0, 
\end{eqnarray}
so that all fictitious walls are found in the region $x>0$  (see Fig.\ref{Fig:1}c). 
Next we use Eq.(\ref{c3}) to analyse reflection in two spatial dimensions. 
\begin{figure}[h]
\includegraphics[angle=0,width=0.48\textwidth]{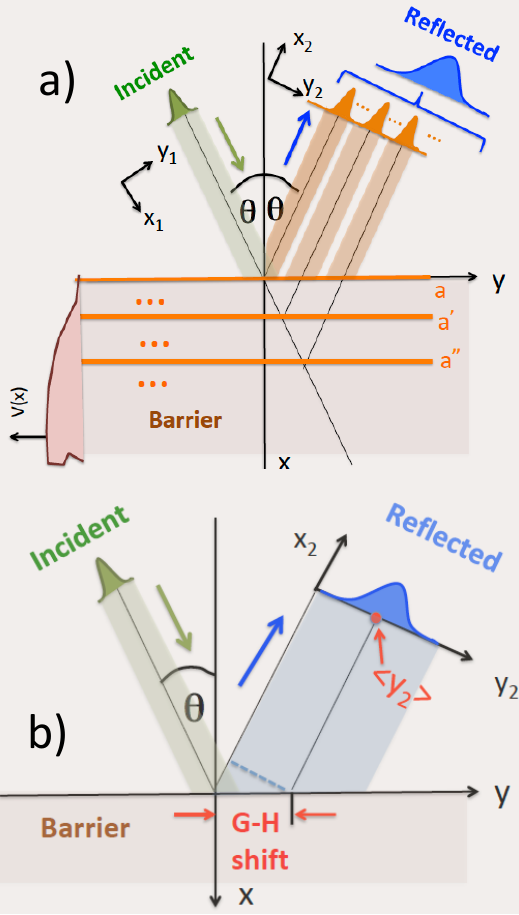}
\caption {a) In two dimensions,  a Gaussian beam is incident on a potential barrier $V(x,y)=V(x) \theta (x)$.
The reflected beam is a superposition of all beams, reflected by a variety of infinite walls placed at $x>0$.
b) The centre of the reflected beam [cf. Eq.(\ref{k5a})] and the Goos-H\"anchen shift (\ref{k6}).} 
\label{Fig:5}
\end{figure}
\section{Quantum Goos-H\"anchen effect}
In two dimensions [${\bf r}=(x,y)$, ${\k}=(k_x,k_y)$],  consider a potential occupying the $x>0$ half-space 
$V(x,y)=V(x)\theta(x)$. Now if the particle's initial state, localised in the potential-free
region $x<0$,  is given by 
 \begin{eqnarray} \label{k0}
\Psi(\r,0)=\int_{-\infty}^\infty d\k \la \r|\k\ra\la \k|\Psi(0)\ra, \q \la \r|\k\ra =\frac{ \exp(i\k\r)}{2\pi}, \q
\end{eqnarray}
for the reflected WF one finds 
 \begin{eqnarray} \label{k1}
\Psi_R(\r,t)=
\frac{1}{2 \pi}\int_{-\infty}^\infty dk_x \int_{-\infty}^\infty dk_y R(k_x)\times \n 
\exp[-ik_x x+ik_y y-i E(k)t]
 \la \k|\Psi(0)\ra.
\end{eqnarray}
where $R(k)$ is the reflection coefficient for one dimensional potential $V(x)$. 
\newline
Acting as before, for 
 a barrier not supporting bound states we find
  \begin{eqnarray} \label{k2}
\G_R(\r,t ) = \int_{0}^\infty da\, \xi_a (a)\,  \G_R^\infty (\r,t|a) \q [x<0],
\end{eqnarray}
where $ \G^\infty_R (\r,t|a)$ is the freely propagating  initial state, reflected by a hard wall 
placed at $x=a$,  
  \begin{eqnarray} \label{k3}
  \G^\infty_R (x,y,t|a)=-\Psi_0(2a-x,y,t).
\end{eqnarray}
\newline
Let $\Psi_0(x,y,t)$ be a transverse Gaussian beam (or a WP very broad in the direction of propagation), making an angle $\theta$ with the normal to the surface $x=0$  (see Fig. 5a).
In the coordinate system $(x_1,y_1)$, obtained by  rotating  the axes $(x,y)$  counter clockwise by $\theta$, we have
  \begin{eqnarray} \label{k4}
\Psi_0(x_1,y_1,t)=G_0(y_1)\exp(i p x_1-ip^2t/2m),
\end{eqnarray}
where we neglect the beam's broadening by putting,
 \begin{eqnarray} \label{k4a}
G_0(y)=\left (\frac{2}{\pi\Delta y^2 }\right)^{1/4}\exp\left ( -\frac{y^2}{\Delta y^2}\right).
\end{eqnarray}
(see Appendix G).
Returning to the spatial shifts (\ref{e5}), $x'=2a$, 
after a simple exercise in trigonometry, 
we find  ($x<0$)
  \begin{eqnarray} \label{k5}
\G_R(x_2, y_2, t ){\xrightarrow[ x_2 \to \infty ] {}}\exp[ipx_2-iE(p)t] \times \n
\int_0^\infty dx' G_0(y_2-x'\sin \theta)\exp(ip_xx')\xi(x')
\end{eqnarray}
where the  axes $(x_2,y_2)$ are obtained by reflecting  the axes  $(x_1,y_1)$ about the plane $x=0$ (see Fig.\ref{Fig:5}a).] 
Again we have a kind of quantum measurement. Detecting the particle at some $y_2$ one can deduce
that it has experienced a spatial shift in the interval $(y_2 -\Delta y_2)/\sin \theta \lesssim x'\lesssim (y_2 +\Delta y_2)/\sin \theta$.
The measurement of $x'$ is the more accurate the smaller is the ratio between the width of the beam $\Delta y_2$ and the sine 
of the angle of incidence $\theta$, $\Delta x' =\Delta y_2/\sin \theta$. 
\newline
The quantity  of interest is the  Goos-Hänchen shift $\delta y_{\text{GH}}$, i.e., the distance by which the reflected beam is displaced relative to the same 
beam reflected by a hard wall placed at $x=0$ (see Fig.\ref{Fig:5}b). We note that superposition of the Gaussians in Eq.(\ref{k5}) does not have to be
 a Gaussian itself. However, one can always calculate the centre of the beam, the way  it was done in Eq.(\ref{f6})
  \begin{eqnarray} \label{k5a}
\la y_2 \ra= \frac{\int_0^{\infty} dy_2\, y_2\, |\G_R(x_2\to \infty, y_2, t )|^2}{ \int_0^{\infty} dy_2\, |\G_R(x_2\to \infty, y_2, t )|^2},
\end{eqnarray}
and from it obtain
 \begin{eqnarray} \label{k6}
S_{\text{GH}} =\frac{\la y_2 \ra}{\cos (\theta)}.
\end{eqnarray}
In the limits of a narrow and broad beam, we find 
expressions similar to Eqs.(\ref{g3}) and (\ref{h2}), respectively, 
 \begin{eqnarray} \label{k6a}
S_{\text{GH}}{\xrightarrow[ \Delta y/\sin \theta \to 0 ] {}}  \tan(\theta)\frac{\int_0^\infty dx' x' |\xi(x')|^2}{\int_0^\infty dx' |\xi(x')|^2}
\end{eqnarray}
and 
 \begin{eqnarray} \label{k7}
S_{\text{GH}} {\xrightarrow[ \Delta y/\sin \theta \to \infty  ] {}} \tan(\theta)
\R\left[
\frac{\int_0^\infty dx' x' \exp(ip_x x' )\xi(x')}{\int_0^\infty dx' \exp(ip_xx')\xi(x')}
\right ] \q\q \n
= \tan(\theta) \frac{\partial \phi_R(k)}{\partial k}
{\biggr\rvert}_{k=p_x=p\cos\theta}\q\q\q\q
\end{eqnarray}
To calculate $\la y_2 \ra$ for any $\Delta y$, we write $G_0(y)=\int \tilde G_0(q) \exp(iyq)dq$
and use Eqs. (\ref{b3}) and  (\ref{ae4}). The result is
\begin{eqnarray} \label{k8}
\la y_2\ra\equiv 
\I\left [\frac{ \int_{-\infty}^\infty dq F^*(q,p_x, \theta)\partial_q F(q,p_x, \theta)}{\int_{-\infty}^\infty dq |F(q,p_x, \theta)|^2}\right ], \n
F(q,p_x, \theta)\equiv R(p_x-q\sin\theta)\tilde G_0(q). \q\q
\end{eqnarray}
Finally, in the paraxial approximation of Appendix G, effects of beam's broadening can be included by replacing $G_0(y)$ and $\tilde G_0(q)$ 
in Eqs.(\ref{k4}) and (\ref{k8}), by their coordinate-dependents versions (\ref{ax4}) and (\ref{ax6}). 
This approach is also useful beyond the paraxial approximation. For example, consider a doubly Gaussian wave packet,  reflected by potential step (see Section VIII) in the regime of total reflection (i.e., when all energies in the wave packet are much lower than the step height).  In this case, both the longitudinal delay and the lateral displacement can be directly related to the average depth at which the particle is reflected,
\begin{equation}
\langle x_2 \rangle=-x_0 - \frac{p_0}{m} t+\langle x \rangle \cos\theta, \q\q \langle y_2 \rangle=\langle x \rangle \sin\theta. 
\end{equation}
The  average depth at which the reflection occurs is related to the derivative of the phase [see (\ref{k0YC})] of the reflection coefficient (\ref{m2YC})
\begin{eqnarray}
\langle x \rangle \approx \iint dk_{x_2} dk_{y_2} {\frac{d \phi_R(k_x)}{dk_x}\bigg\vert_{k_x=k_{x_2} \cos\theta - k_{y_2} \sin\theta}}\\
\times\frac{2}{\pi} \frac{\exp(-2 k_{y_2}^2/\Delta k_{y_2}^2)}{\Delta k_{y_2} } \frac{\exp[-2 (k_{x_2}-k_0)^2/\Delta k_{x_2}^2]}{\Delta k_{x_2}}.\nonumber
\end{eqnarray}
Here both  the wave packet widths $\Delta k_{x_2}$ and $\Delta k_{y_2}$ must be sufficiently small, so that adding them to the average momentum $k_0$ keeps the corresponding energies significantly lower than the barrier height.
\section{Conclusions}
Above we attempted to give a most detailed description of the quantum Goos-H\"anchen effect that an elementary theory can offer.
For a classical particle with a known momentum launched at oblique angle at a repulsive potential, 
 the distance $S_{\text{GH}}$ between the exit and entry point at the surface, where the potential vanishes, can be  found 
by multiplying its lateral velocity $v_{\parallel}$ by the duration  $\t$ its trajectory spends in the potential: $S_{\text{GH}} =v_{\parallel}\t$. 
\newline
A straightforward extension of  this analysis to the quantum case is, however, not possible.
In one dimension, the quantum state of a particle with a known momentum $p=mv$  is a plane wave, and the moments 
when it enters and leaves the potential region simply cannot be defined. 
One notes, however, that the reflected state is a superposition (\ref{e5}) of multiple plane waves 
with phases corresponding to spatial  displacements, or shifts, $x'$, induced by the potential. 
One notes also that in experiment where one prepares the particle in a wave packet state with
a mean momentum $p$, and then observes it at $x$,  the shifts are restricted to
the range, determined by the WP's coordinate width $\Delta x$. In the classical limit of this \e{quantum measurement}, 
the unique displacement $x'_s$, with which the reflected particle leaves the potential, corresponds to the stationary region 
where rapid oscillations of the amplitude distribution $\xi(x')$ slow down. This classical value allows one to deduce the duration $\t$ that the particle's
trajectory spends in the  potential as $x'_s/v$. 
\newline
There are two good reasons why one cannot continue to use the time argument beyond the classical limit, e.g., by converting 
each spatial shift into a duration using $\t(x')=x'/v$. One is that some of these \e{durations} will have to be negative 
in the case of a potential well, or, indeed, of any potential supporting bound states. The idea that a particle 
which makes a U-turn before entering the potential spends in the scatterer  a \e{negative duration}
does not help explain the phenomenon in classical-like terms. 
[We are aware of the recently reported  \e{experimental evidence of negative times} \cite{St},
and remain sceptical.]  The notion that a particle's state can acquire a negative phase shift
can, however,  be made without causing a further controversy.
\newline
The other reason is that, as with any quantum measurement of this kind, one faces a dilemma rooted in the Uncertainty
Principle \cite{FeynL}. One can determine the shift experienced by the particle accurately, but the reflection probability will 
not correspond to that of a particle with a momentum $p$. In fact, it will vanish. 
Or, one can keep the probability (almost) intact, but loose all the knowledge about the $x'$s to interference. 
In this case it is still possible to find the displacement of the centre of mass of a very broad reflected 
WP and, from it the \e{phase time}  (\ref{h2}). However, both are  related to the first moment of an oscillatory
complex valued distribution, and can be negative even for a potential barrier, 
where all shifts $x'$s are positive
(for an example see \cite{Neg}). 
\newline
One can represent the scattering potential by a superposition of virtual hard walls, 
distributed in such a manner that the outgoing WP is a sum of all WPs reflected by each of wall. 
Such a representation is not without its problems, since in the case of a well, some of the fictitious 
walls would need to be placed in the potential-free region, implying that the scattering occurs 
even before the particle reaches the scatterer. Still the idea is somewhat less confusing than the one invoking negative durations, 
and can be used for illustrative purposes.
\newline
For example, the hard wall picture helps one to visualise what happens in the quantum Goos-Hänchen case \cite{GH1}.
The outgoing Gaussian beam is composed of the beams reflected by each fictitious wall (see Fig.\ref{Fig:5}a).
The cross section of the beam's wave function is a superposition of Gaussians (we neglect the lateral 
spreading of the wavepacket during propagation) similar to one would obtain in a von Neumann quantum measurement.  
As the width of the incident beam increases (the measurement becomes less accurate)
the Goos-H\"anchen shift changes from Eq.(\ref{k6a}) to Eq.(\ref{k7}), involving
the \e{strong} and the \e{weak} values of the coordinate shift $x'$.
\newline

\section*{Appendix A. Scattering by a one-dimensional barrier}
Consider scattering by a one-dimensional potential barrier $V(x)$, contained  between $x=0$ and $x=d$, 
\begin{eqnarray} \label{-1}
V(x)\equiv 0, \q \mathrm{for} \q  x<0 \q \mathrm{and} \q x>d.
\end{eqnarray}
The left (right) scattering states, for a particle incident from the left (right), 
\begin{align}\label{0}
\la{x}|\Phi_l(k)\ra =\frac{1}{\sqrt{2\pi}} \left[  e^{ikx} + R_l(k)e^{-ikx}  \right] ,\q & x<0\\
 =  \frac{1}{\sqrt{2\pi}}  T_l(k)e^{ikx}, \q & x>d \nonumber \\
\la{x}\ket{\Phi_r(k)} =\frac{1}{\sqrt{2\pi}} \left[  e^{-ikx} + R_r(k)e^{ikx}  \right] ,\q & x>d \nonumber \\
 =  \frac{1}{\sqrt{2\pi}}  T_r( k)e^{-ikx},\q &  x<0 \nonumber 
\end{align}
form a complete orthonormal basis, 
\begin{eqnarray} \label{-2}
\int_{0}^\infty dk\Big [|\Phi_l(k)\ra\la \Phi_l(k)| +|\Phi_r(k)\ra\la \Phi_r(k)|\Big ] =1.
\end{eqnarray}
It follows from the Wronskian relations between the states (\ref{0})
and the fact that the complex conjugate $\Phi^*_{l,r}(k,x)$ are also solutions of the Schr\"odinger equation (SE), that
\begin{align} \label{1}
T_l(k)=T_r(k)=T(k), \q &R_l(k)T^*(k)= -R^*_r(k)T(k),\\
T(-k^*)=T(k)^*, \q &R_{l,r}(-k^*)= R^*_{l,r}(k).\nonumber
\end{align}
The initial state is a wave packet $|\Psi(x,0)\ra$ completely localised to the left of the barrier, which we expand in plane waves $|k\ra$ as 
\begin{align} \label{2a}
|\Psi(x,0)\ra = \int_{-\infty}^\infty dk |k\ra\la k|\Psi(0)\ra\equiv  \int_{-\infty}^\infty dk A(k)|k\ra, \n \la x|k\ra=\frac{1}{\sqrt{2\pi}} e^{ikx}, 
\end{align}
To evolve the state in time we expand it in the scattering states (\ref{0}),
\begin{align} \label{2}
\G(x,t) =& \int_{0}^\infty dk \la \Phi_r (k)|\Psi(0)\ra e^{-iE(k)t} \la x|\Phi_r(k)\ra +\\
&\int_{0}^\infty dk \la \Phi_l( k)|\Psi(0)\ra e^{-iE(k)t} \la x|\Phi_l(k)\ra ,
\end{align}
where $E(k)=k^2/2m$, and $m$ is the particle's mass.
Using $ \int_{-\infty}^\infty dk^\prime \int_{-\infty}^\infty dx' |x'\ra \la x'| k'\ra \la k'|=1$ we find
\begin{eqnarray} \label{3}
\la \Phi_{l,r}(k)|\Psi(0)\ra=\int_{-\infty}^\infty dk^\prime \int_{-\infty}^\infty dx'\n
\la \Phi_{l,r}(k)|x'\ra\la x'|k'\ra \la k'|\Psi(0)\ra
\end{eqnarray}
Now, since the initial state lies to the left of the barrier, i.e., at $x<0$, we can use the corresponding expressions for $\Phi^*_{l,r}(k,x)$ [cf. Eq.(\ref{0})]
and extend integration over $x'$ to $(-\infty, \infty)$. With $(2\pi)^{-1}\int_{-\infty}^\infty\exp(ikx)\, dx=\delta(k)$, this gives 
\begin{align} \label{4}
\la \Phi_{l}(k)|\Psi(0)\ra=(2\pi)^{-1}\int_{-\infty}^\infty dk^\prime \int_{-\infty}^\infty dx'&\\
\left[ e^{-ikx'} + R_{l}^*(k)e^{ikx'} \right]e^{ik'x'}\la k'|\Psi(0)\ra =\nonumber &\\
 \la k|\Psi(0)\ra +R_l^*(k) \la -k|\Psi(0)\ra,\nonumber &\\
\la \Phi_{r}(k)|\Psi(0)\ra=(2\pi)^{-1}\int_{-\infty}^\infty dk^\prime \int_{-\infty}^\infty dx' \nonumber &\\
T^*(k)e^{ikx'}e^{ik'x'}\la k'|\Psi(0)\ra =T^*(k) \la -k|\Psi(0)\ra.\nonumber &%
\end{align}
Inserting (\ref{4}) into (\ref{2}) for  {{$x>d$}} yields
\begin{align} \label{5}
\G&(x,t) =\frac{1}{\sqrt{2\pi}}\times\\
 \Big\{ &\int_{0}^\infty dk
  [\la k|\Psi(0)\ra +R_{l}^*(k) \la -k|\Psi(0)\ra] T(k) e^{ikx-iE(k)t}\nonumber\\
 +&\int_{0}^\infty dk T^*(k) \la -k|\Psi(0)\ra[e^{-ikx} + R_r(k) e^{ikx}] e^{-iE(k)t}\Big\},\nonumber
\end{align}
 which, after taking into account (\ref{1}), and changing $k\to -k$  simplifies to
 \begin{eqnarray} \label{6}
\G(x,t) = \frac{1}{\sqrt{2\pi}}\int_{-\infty}^\infty dk  T(k)  \la k|\Psi(0)\ra e^{ikx-iE(k)t}.
\end{eqnarray}
 Similarly, for {\boldmath{$x<0$}} we have
 \begin{eqnarray} \label{7}
\G(x,p,t) = \int_{0}^\infty dk [\la k|\Psi(0)\ra +R_{l}^*(k) \la -k|\Psi(0)\ra]\n
 \times \frac{1}{\sqrt{2\pi}}[e^{ikx} + R_l(k)e^{-ikx}] e^{-iE(k)t}\q\q\q\q\n
 +\frac{1}{\sqrt{2\pi}}\int_{0}^\infty dk T^*(k) \la -k|\Psi(0)\ra T(k)  e^{-ikx} e^{-iE(k)t},
\end{eqnarray}
 Or, recalling that $|T(k)|^2+|R_l(k)|^2=1$ and $R_l^*(k)=R_l(-k)$,
  \begin{eqnarray} \label{8}
\G(x,t) =\frac{1}{\sqrt{2\pi}} \int_{-\infty}^\infty dk \la k|\Psi(0)\ra e^{ikx-iE(k)t} +\n
 \frac{1}{\sqrt{2\pi}} \int_{-\infty}^\infty R_l(k) dk \la k|\Psi(0)\ra e^{-ikx-iE(k)t}.
\end{eqnarray}
Validity of (\ref{8}) is easily proven. Define
  \begin{eqnarray} \label{8a}
I_1= \q \int_{0}^\infty dk \la k|\Psi(0)\ra e^{ikx -iE(k)t} 
\end{eqnarray}
 \begin{eqnarray} \label{8b}
 I_2=\int_{0}^\infty dk R_{l}^*(k) \la -k|\Psi(0)\ra e^{ikx-iE(k)t}+\n
  \int_{0}^\infty dk R_{l}(k) \la k|\Psi(0)\ra e^{-ikx-iE(k)t}=\n
\int_{-\infty}^\infty dk R_{l}(k) \la k|\Psi(0)\ra e^{-ikx-iE(k)t}
\end{eqnarray}
 \begin{eqnarray} \label{8c}
I_3= \int_{0}^\infty dk |R_{l}(k)|^2 \la -k|\Psi(0)\ra e^{-ikx-iE(k)t}+\n
\int_{0}^\infty dk |T(k)|^2 \la -k|\Psi(0)\ra  e^{-ikx-iE(k)t}= \n
 \int_{0}^\infty dk \la -k|\Psi(0)\ra e^{-ikx-iE(k)t}=\n 
 \int_{-\infty}^0 dk \la k|\Psi(0)\ra e^{ikx-iE(k)t} 
  \end{eqnarray}
Then adding $I_1+I_2+I_3$ yields $\int_{-\infty}^\infty dk \la k|\Psi(0)\ra e^{ikx-iE(k)t}$
\newline
Equations (\ref{6}) and (\ref{8}) allow one to monitor the time evolution of the wave function in the regions $x<0$ and $x>d$, 
but not inside the potential, $0<x<d$, where one needs to know the behaviour of $\Phi_{l.r}(k,x)$,  not specified by Eqs.(\ref{0}). 
For example, the initial WP located to the left of the potential, is represented by the first term in Eq.(\ref{8}). As time progresses, 
it gradually  \e{disappears from the screen}, and is replaced by the second term in (\ref{8}) in the region $x<0$, while 
the transmitted part (\ref{6}) appears in the  $x>0$ region. Note that (\ref{6}) and (\ref{7}) are valid even when the WP expansion 
(\ref{2aC}) contains negative momenta, propagating away from the barrier, as we illustrate next. 
\section*{Appendix B. Gaussian wave packets}
It is common practice to study scattering of Gaussian states.
Choosing 
\begin{eqnarray} \label{1a}
\la k|\Psi(0)\ra= \left ({2\pi^{-1} \Delta k^2}\right )^{-1/4}\exp\left [-\frac{(k-p)^2}{\Delta k^2}- ikx_0\right ]\q
\end{eqnarray}
with $x_0<0$ and $|x_0| >> 2/\Delta k$ yields an initial Gaussian wave packet, set a distant $|x_0|$
to the left of the barrier, and moving with velocity $v=p/m$ towards it or away from it, depending on the 
sign of $p$
\begin{eqnarray} \label{2aC}
\la x|\Psi(0)\ra=\left (\frac{\pi \Delta x^2}{2}\right )^{-1/4} \exp\left[ -\frac{(x-x_0)^2}{\Delta x^2} + ip(x-x_0)\right ]\q\q
\end{eqnarray}
where $\Delta x=2/\Delta k$.
For $p < \Delta k$, the initial WP contains also negative momenta, even though it moves towards the barrier.
Similarly, with $p <0$, the WP moving now {\it away} from the barrier contains positive momenta. 
The first term in Eq.(\ref{8}), describing free motion of the initial wave packet, is now given by 
\begin{align} \label{3a}
\Psi(x,t)&= \int_{-\infty}^\infty dk \la k|\Psi(0)\ra
 e^{i k x-iE(k)t}\n &=e^{ip(x-x_0)-iE(p)t}G_0(x-pt/m, t) 
\end{align}
where the  envelope $G$, which contains the effects of spreading is given by 
\begin{eqnarray} \label{3ab}
G_0\left(x-\frac{pt}{m}, t\right) =\left (\frac{2\Delta x^2}{\pi\Delta x_t^4 }\right )^{1/4}&\exp\left [ -\frac{(x-vt -x_0)^2}{\Delta x_t^2}\right],\\
 &\Delta x_t\equiv \sqrt{\Delta x^2+\frac{2it}{m}},\nonumber
\end{eqnarray}
If the spreading can be neglected, e.g., for $2t/m << \Delta x^2$,  the envelope travels without distortion with a velocity $v=p/m$, 
\begin{eqnarray} \label{5a}
G_0(x-\frac{pt}{m}, t) \approx \left (\frac{2}{\pi\Delta x^2 }\right )^{1/4}\exp\left [ -\frac{(x-vt -x_0)^2}{\Delta x^2}\right].\q\q
\end{eqnarray}
Equations (\ref{3a})-(\ref{5a}) are valid, we repeat, even  if the initial Gaussian state (\ref{1a}) contains momenta in the direction opposite to the WP's velocity $v$. 
\section*{Appendix C. Free motion and reflection from an infinite (hard) wall}
To check the consistency of Eqs.(\ref{6}) and (\ref{8}), we consider two simple cases. 
For free motion, $V(x) \equiv 0$, $T(k)=1$, $R(k)=0$ for $-\infty<k< \infty$, Eq.(\ref{6}) recovers, as expected, the correct answer (\ref{3a}) for $x>d$.
\newline 
More interesting is the case of reflection by an infinite wall placed at some $a>0$, $V(x)=0$ for $x< a$, and $V(x)=\infty$ for $x\ge a$. 
Now we have
\begin{align} \label{1ab}
R(k)&= -\exp(2ika), \q -\infty<k< \infty,\\
\xi(x')&=-\delta (x'-2a),\nonumber
\end{align}
where the negative sign ensures that the wave function vanishes at $x=a$. 
Now the reflected WP entering the region $x<0$ is, apart from the change of the sign, a copy of the freely propagating initial WP, 
reflected about $x=a$. 
\begin{align} \label{2ab}
\Psi_R(x,t)&=
 \frac{1}{\sqrt{2\pi}} \int_{-\infty}^\infty dk  R_l(k) \la k|\Psi(0)\ra e^{-ikx-iE(k)t}\\
   & =-\frac{1}{\sqrt{2\pi}}\int_{-\infty}^\infty dk \la k|\Psi(0)\ra e^{-ik(x-2a)-iE(k)t}.\nonumber
\end{align}
This is also the correct result which we will need in the following. 
\section*{Appendix D. Quantum measurements}
Consider, in one dimension, a particle (pointer) (P) with position $f$ and momentum $\p$, briefly interacting with a quantum system (S), 
\begin{eqnarray} \label{ac1}
\h=\h^S+\h^P+\h_{int},  
\end{eqnarray}
where 
\begin{eqnarray} \label{ac2}
\h^P=\hat \p^2/2\mu, \q
\h_{int}=\hat \p\B\delta(t+\epsilon), \q  \epsilon \to 0,
\end{eqnarray}
 where $\mu$ is the pointer's  mass, and $\B=\sum_i|b_i\ra B_i\la b_i|$ is the system's operator to be
measured. The system is pre- and post-selected in states $|I\ra$ and $|F\ra$ at $t=0$ and $t$, respectively. 
At $t=0$ the pointer is prepared in the Gaussian state (\ref{2a}) [$x\to f$, $p\to \p$, $\Delta x \to \Delta f$, $x_0 \to f_0=0$], 
so that the composite's initial state is given by $|\Phi(0)\ra=|I\ra|\Psi_0\ra$.
At time $t$ the observer accurately determines the pointer's position, hoping in this way to learn something 
about the value of $\B$ at the time the system interacted with the pointer.
The pointer's state after the system was found in $|F\ra$ at $t$,  is
\begin{eqnarray} \label{ac3}
\la f|\la F|\Phi(t)\ra=\sum_i \Psi(f-B_i,t)\A^S(F\gets b_i\gets I),
\end{eqnarray}
where $ \Psi(f-B_i,t)$ is given by Eqs.(\ref{3a})-(\ref{3ab}), and  $\A_i^S$ is the system's transition amplitude, 
\begin{eqnarray} \label{ac4}
\A^S(F\gets b_i\gets I)\equiv \la F|\exp(-i\h^St)|b_i\ra\la b_i|I\ra.
\end{eqnarray}
It is convenient to introduce an amplitude distribution $\xi(f')$ (if required, the sum  can be replaced by an integral)
\begin{eqnarray} \label{ac5}
\xi(f') \equiv \sum_i \delta(f'-B_i)  \A^S(F\gets b_i\gets I),
\end{eqnarray}
so that (\ref{ac3}) can be rewritten as 
\begin{eqnarray} \label{ac6}
\Phi(f,t) \equiv \la f|\Phi(t)\ra=\int df' \Psi(f-f',t)\xi(f'),
\end{eqnarray}
where $\xi(f')$ is the probability amplitude of the quantity, represented by the operator $\B$, having a value $f'$. 
Suppose we can neglect the spreading of the pointer's wave packet, prepared with mean momentum $\p=0$,
$ \Psi(f-f',t)\approx\left (\frac{2}{\pi\Delta x^2 }\right )^{1/4}\exp\left [ -\frac{(f-f')^2}{\Delta f^2}\right] \equiv G_0(f-f')$. 
Now only the values in the interval $f-\Delta f \lesssim f' \lesssim f+\Delta f $ contribute to $\Phi(f,t)$
and by determining $f$ accurately one measures $\B$ to a (quantum) accuracy $\Delta f$.
Note that as $\Delta f \to \infty$ the probability of the post-selection, $P(F)$, is no longer affected by the pointer, 
$P(F) \to |\la F|\exp(-i\h^S t)|I\ra|^2$. 
\newline
If the spreading can still be neglected, but the pointer is prepared with non-zero mean momentum $\p\ne 0$, 
\begin{eqnarray} \label{ac7}
\Phi(f,t) \equiv \la f|\Phi(t)\ra=\exp[i\lm f - iE(\lm)t]\times \n
\int df' G_0(f-f')\exp (-i\lm f')\xi(f'),
\end{eqnarray} 
and finding the pointer at $f$ one can still know that 
$\B$ had the value roughly  in the interval $(f-\Delta f ,f+\Delta f )$. However, now sending $\Delta f \to \infty$
does not fully decouple the pointer from the system, as $P(F) \to |\int df' \exp(-i\p f')\xi(f')|^2\ne  |\la F|\exp(-i\h^S t)|I\ra$.
The similarity between Eqs.(\ref{ac6}) and (\ref{b4}) provides the required analogy. 
\section*{Appendix E.  Rectangular barrier and potential step}
Consider  a rectangular barrier $V(x)$  of  height $V=const$, occupying the region $0\le x \le d$. 
For a particle of mass $m$, the reflection amplitude is well known to be
 \begin{eqnarray} \label{ad1}
R(k)=\frac{(k^2-q^2)\sin qd}{ 2ikq\cos qd+(k^2+q^2)\sin qd} 
\end{eqnarray}
where $q(k)\equiv \sqrt{k^2-2mV}\equiv  \sqrt{k^2-W^2} $. The amplitude is an entire function in the complex $k$-plane, unchanged by replacement $q\to -q$ and satisfying, 
for real $k$ , $R(-k)=R^*(k)$. 
{For finite $ d $},
the reflection coefficient has zeros in the real $ k $ axis and
{isolated} poles in the lower half plane [complex zeroes of the denominator in Eq.(\ref{ad1})].
The zeros at $ k_l=\pm\sqrt{(l\pi/d)^2+W^2} $ correspond to energies 
at which the barrier is perfectly transparent.
The poles, corresponding to the zeros of the denominator in Eq.(\ref{ad1}),
 are associated with the resonances {(or quasi-stationary states) \cite{BZP}} at the energies above 
the barrier, $k^2/2m> V$.
\begin{figure}[h]
	\centering\includegraphics[angle=0,width=.48\textwidth]{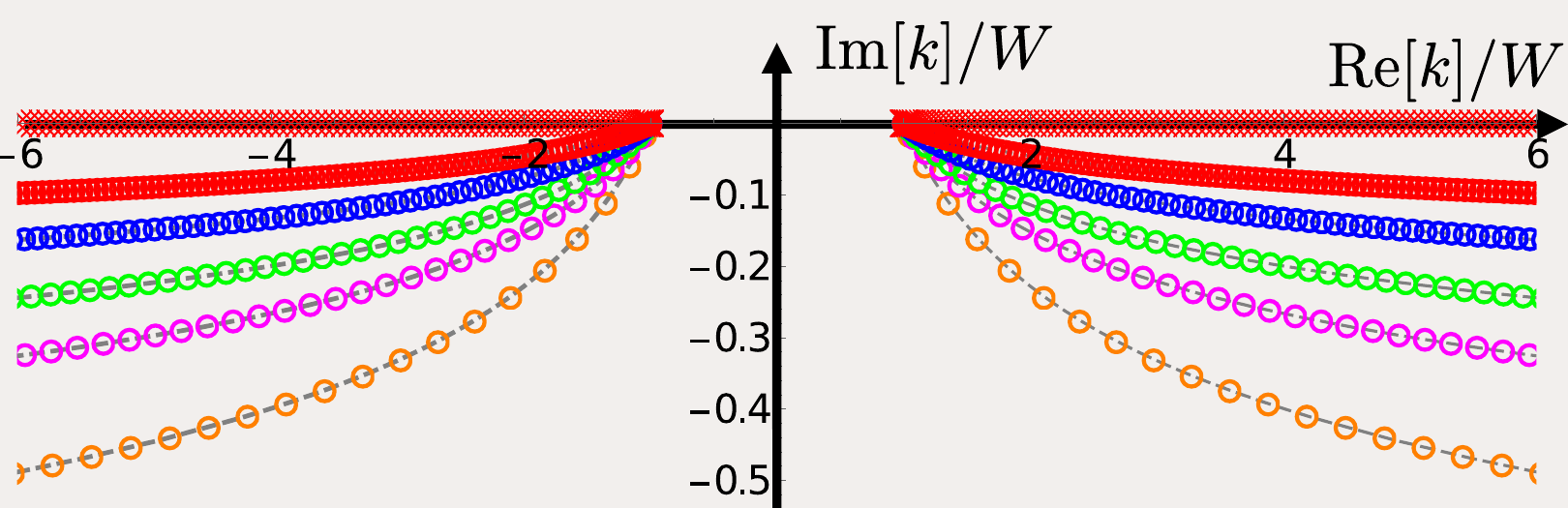}
	\caption {The exact (numerical) poles of Eq.(\ref{ad1}) evaluated for 
		$ d=10,15,20,30,50 $ (circles) in
				orange, magenta, green, blue and red respectively. 
				Also shown the zeros of $ R(k) $ for $ d=50 $ (red crosses).
		The dashed grey lines show the approximations from Eq.\eqref{eq:approximation}.
	}
	\label{Fig:poles}
\end{figure}
For large $ d $, their approximate position in the $ k $-plane is given by, 
\begin{align}\label{eq:approximation}
k_n&\approx \chi_n(d)-
\frac{1}{d}\thinspace
\frac{\beta_n(d)}{ \chi_n(d)}\times
\bigg(
i\log\left|{\frac{\chi_n(d)+\beta_n(d)}{\chi_n(d)-\beta_n(d)}}\right|\\
&-\arg\left(\frac{\chi_n(d)+\beta_n(d)}{\chi_n(d)-\beta_n(d)}\right)\bigg)
+\mathcal{O}(d^{-2}),\nonumber
\end{align}
where $ \beta_n(d)\equiv n\pi/d $ and $ \chi_n(d)\equiv\sqrt{\beta_n(d)+W^2} $  (see Fig.\ref{Fig:poles}).
As $ d\to\infty $ the poles approach the real axis and together 
with the zeros form cuts drawn from the branch points 
$ k=\pm \sqrt{W} $ to $ \pm\infty $.
A similar behaviour of densely coalescing poles was reported, e.g. in
\cite{Dense_poles}.
This branch cut corresponds
to the one present in the reflection amplitude for the step potential found in Eq.\eqref{ad2}.
Thus, to  calculate $\xi (x')=(2\pi)^{-1}\int _{-\infty}^\infty dk \exp(-ikx') R(k)$ we may shift the contour of integration into the upper half-plane by 
putting $k=k+i\epsilon$. $\epsilon \to 0$, so $\xi(x')=0$ for $x'<0$.
Note that on the contour shown in Fig.\ref{Fig:6}, $q(-k)=q(k)$ for $|k|\le W$, and $q(-k)=-q(k)$ for $|k| > W$.

To obtain the reflection amplitude for a potential step, $V(x) = V\theta(x)$, we send $d\to\infty$ before taking the limit $\epsilon \to 0$, 
and retain only the growing exponentials. (For example, for $k>W$, we have $q(k+i\ep) \approx q(k) +ik\ep/q$ so that
$\exp[-iq(k+i\ep)d]\to \infty$ as $d\to \infty$.) This gives the well known expression
 \begin{eqnarray} \label{ad2}
R(k,q(k))=\frac{k-q}{k+q}, 
\end{eqnarray}
which has no poles, but two branch cuts from $-\infty$ to $-W$, and from $W$ to $\infty$, formed by the coalescence of the resonance poles
as $d$ increases. To evaluate the amplitude distribution of the shifts, $\xi(x')$, we integrate along the contour passing just above the cuts (see Fig.\ref{Fig:6}).
For $x'<0$ the contour can be closed in the upper half-plane where $R(k)$ has no singularities, so $\xi(x')\equiv 0$. 
\begin{figure}[h]
\includegraphics[angle=0,width=.48\textwidth]{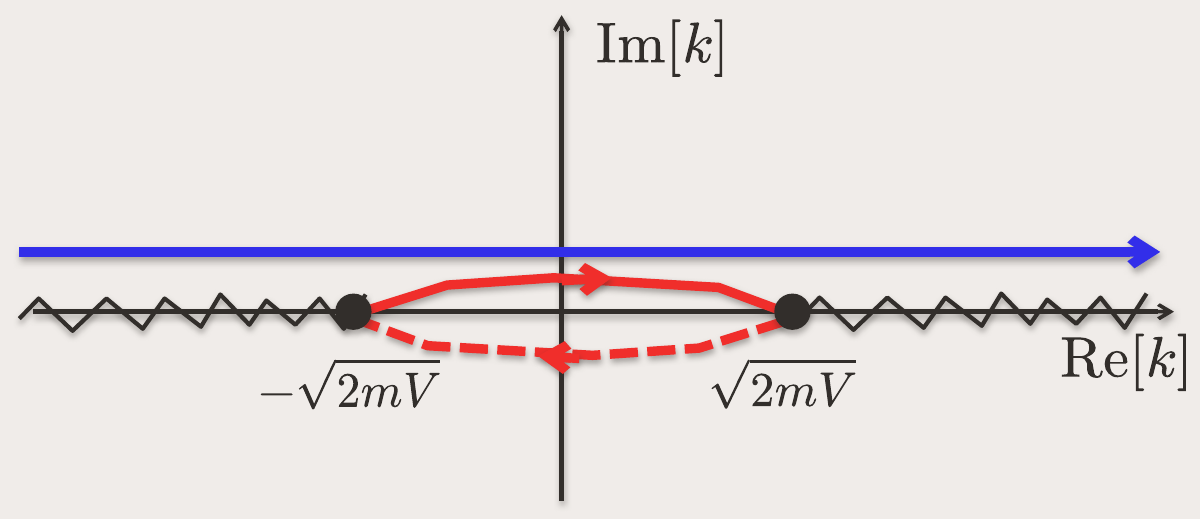}
\caption {For $x'>0$, the contour of integration running just above the real $k$-axis
can be replaced by a loop around the branching points at $k=\pm \sqrt{2mV}$}
\label{Fig:6}
\end{figure}
For $x'>0$, the contour can be transformed into a loop around the branch points as shown in Fig.\ref{Fig:6}, and we have
  \begin{align} \label{ad3}
\xi(x'\ge 0) &= (2\pi)^{-1}\oint dk R(k)\exp(-ikx')\\
&=-\frac{4W}{\pi}\int_{0}^{1}dt t\sqrt{1-t^2}\sin(Wx't),\nonumber\\
&=\frac{2}{\sqrt {2mV}}\frac{d}{dx'}\left (  \frac{J_1(\sqrt{2mV}x')}{x'}\right ) ,\nonumber
\end{align}
and $\xi(x'<0)\equiv 0$.
In (\ref{ad3})  $J_1(z)$ is the Bessel function of the first order \cite{Abram}.
It is easy to check that $\int_0^\infty \xi(x')= R(p=0)=-1$.
Finally, using Eqs. 9.1.30 of \cite{Abram},  one obtains
  \begin{eqnarray} \label{ad3a}
\xi(x'\ge 0) =
 - \frac{2}{x'}\, J_2(\sqrt{2mV} x').
  \end{eqnarray}
\section*{Appendix F. Centre-of-mass of the reflected wave packet }
Calculation of the average 
\begin{eqnarray} \label{ae1}
\la x\ra\equiv \frac{\int_{-\infty}^\infty dx\, x\, |\G_R(x,t)|^2}{\int_{-\infty}^\infty dx\, |\G_R(x,t)|^2},
\end{eqnarray}
where 
\begin{eqnarray} \label{ae2}
\G_R(x,t)=\int_{-\infty}^\infty dk  F(k,t) \exp [-ikx], \n
F(k,t)\equiv R_l(k) \la k|\Psi(0)\ra \exp [-iE(k)t] 
\end{eqnarray}
reduces to evaluation of simple quadratures.
Insertion of the relations
\begin{eqnarray} \label{ae3}
\int dx \exp[i(k-k')x]=2\pi\delta(k-k'),\q\q\q\q \n
\int dx\, x\,\exp[i(k-k')x]=i\pi(\partial_{k'}-\partial_{k})\delta(k-k'). 
\end{eqnarray}
 into (\ref{ae1})-(\ref{ae2}) yields Eq.(\ref{k8}),
\begin{eqnarray} \label{ae4}
\la x\ra\equiv 
\frac{\I \left [\int_{-\infty}^\infty dk F^*(k,t)\partial_k F(k,t)\right ]}{\int_{-\infty}^\infty dk |F(k,t)|^2}.
\end{eqnarray}
\section*{Appendix G. Gaussian beam in paraxial approximation} 
Consider a monoenergetic two dimensional Gaussian beam, propagating freely along the $x$-axis, 
\begin{eqnarray} \label{ax1}
\Psi_0(x,y,t) =\exp(-ip^2t/2m)\times,\q\q\q\q\n
\int dq \tilde G_0(q)\exp[iqy+ i\sqrt{p^2-q^2}(x-x_0)],\q
\end{eqnarray}
where 
\begin{eqnarray} \label{ax2}
\tilde G_0(q)= ({2\pi^{-1} \Delta q^2} )^{-1/4}\exp(-q^2/\Delta q^2). 
\end{eqnarray}
If the spread of the transverse moment $\Delta q$
 is small compared to $p$, 
for $x> x_0$ we have
\begin{eqnarray} \label{ax3}
\Psi_0(x,y,t) =\exp(ip(x-x_0)-ip^2t/2m)\times
\n
\int dq \tilde G_0(q,T)\exp(iqy)
\end{eqnarray}
here $T=T(x)=(x-x_0)/v$ is the time it takes particle with velocity $v=p/m$ to travel a distance $x-x_0$, and 
\begin{eqnarray} \label{ax4}
\tilde G_0(q,T)=({2\pi^3 \Delta q^2} )^{-1/4}\exp\left (-\frac{q^2}{\Delta q^2}-i \frac{ q^2 T}{2m}\right )
\end{eqnarray}
Integration over $dq$ the yields the paraxial approximation for the beam
\begin{eqnarray} \label{ax5}
\Psi_0(x,y,t) =\exp(ipx-ip^2t/2m)G_0(y, T(x))
\end{eqnarray}
where the envelope 
\begin{eqnarray} \label{ax6}
G_0(y, T) =\left (\frac{2\Delta y^2}{\pi\Delta y_T^4 }\right )^{1/4}\exp\left [ \frac{-y^2}{\Delta y_T^2}\right], \q\q\n
 \Delta y_T\equiv \sqrt{\Delta y^2+\frac{2iT}{m}}. \q\q\q\q\q\q\q
\end{eqnarray}
now accounts for beam's broadening. Neglecting the broadening yields  Eq.(\ref{k4}).
\section*{Acknowledgements}
DS acknowledges financial support by the Grant PID2021-126273NB-I00 funded by MICINN/AEI/10.13039/501100011033 and by \e{ERDF A way of making Europe}, as well as by the Basque Government Grant No. IT1470-22. YC is a research associate of the Fund for Scientific Research (F.R.S.-FNRS). YC acknowledges the support of the EU from the EIC Pathfinder Challenges 2022 call through the Research Grant 101115149 (project ARTEMIS). The content reflects only the authors’ views, and the European Commission is not responsible for any use that may be made of the information it contains.

\begin{thebibliography}{99}
\bibitem{FOOT} Note that the initial WP must be prepared at $x_0 << 1/\Om$ is order to avoid overlap 
with the bound state at $E=-\Om$, which decays as $\exp(-|\Om|x)$.
\end{thebibliography}
\end{document}